\documentclass{aastex631} %[linenumbers]

\shorttitle{AASTeX v6.3.1 Sample article}
\shortauthors{Zhang et al.}

\graphicspath{{./}{figures/}}

\begin{document}

\title{Sunspot shearing and sudden retraction motion associated with the 2013 August 17 M3.3 Flare \footnote{Released on May 10, 2022}}

\correspondingauthor{Yanjie Zhang}
\email{zhangyj@pmo.ac.cn}
\correspondingauthor{Zhe Xu}
\email{xuzhe6249@ynao.ac.cn}

\author[0000-0003-1979-9863]{Yanjie Zhang}
\affiliation{Purple Mountain Observatory, Chinese Academy of Sciences, \\
Nanjing 210023, People's Republic of China}
\affiliation{School of Astronomy and Space Science,
University of Science and Technology of China, \\
Hefei, 230026, People's Republic of China}

\author[0000-0002-9121-9686]{Zhe Xu}
\affiliation{Yunnan Observatories, Chinese Academy of Sciences, \\
396 Yangfangwang, Guandu District, Kunming 650216, People’s Republic of China}

\author[0000-0003-4078-2265]{Qingmin Zhang}
\affiliation{Purple Mountain Observatory, Chinese Academy of Sciences, \\
Nanjing 210023, People's Republic of China}
\affiliation{School of Astronomy and Space Science,
University of Science and Technology of China, \\
Hefei, 230026, People's Republic of China}

\author[0000-0003-4787-5026]{Jun Dai}
\affiliation{Purple Mountain Observatory, Chinese Academy of Sciences, \\
Nanjing 210023, People's Republic of China}
\affiliation{School of Astronomy and Space Science,
University of Science and Technology of China, \\
Hefei, 230026, People's Republic of China}

\author[0000-0002-5898-2284]{Haisheng Ji}
\affiliation{Purple Mountain Observatory, Chinese Academy of Sciences, \\
Nanjing 210023, People's Republic of China}
\affiliation{School of Astronomy and Space Science,
University of Science and Technology of China, \\
Hefei, 230026, People's Republic of China}

\begin{abstract}
In this Letter, we give a detailed analysis to the M3.3 class flare that occurred on August 17, 2013 (SOL2013-08-17T18:16).  
It presents a clear picture of mutual magnetic interaction initially from the photosphere to the corona via the abrupt rapid shearing motion of a small sunspot before the flare, and then suddenly from the corona back to the photosphere via the sudden retraction motion of the same sunspot during the flare's impulsive phase.  
About 10 hours before the flare, a small sunspot in the active region NOAA 11818 started to move northeast along a magnetic polarity inversion line (PIL), creating a shearing motion that changed the quasi-static state of the active region. 
A filament right above the PIL was activated following the movement of the sunspot and then got partially erupted. The eruption eventually led to the M3.3 flare. 
The sunspot was then suddenly pulled back to the opposite direction upon the flare onset.  
During the backward motion, the Lorentz force underwent a simultaneous impulsive change both in magnitude and direction. 
Its directional change is found to be conformable with the retraction motion.  
The observation provides direct evidence for the role of the shearing motion of the sunspot in powering and triggering the flare. 
It especially confirms that the abrupt motion of a sunspot during a solar flare is the result of a back reaction caused by the reconfiguration of the coronal magnetic field.
\end{abstract}


\section{Introduction} \label{s:intro}
Ever since the discovery made by \citet{Car1859} and \citet{Hod1859}, solar flares have received intensive research for the striking scene of rapid energy release as well as their close relationship with disastrous space weather \citep{priest2002, Fletcher2011, Lin2003}.
It is now confirmed that the energy required for flares comes from the coronal magnetic field directly \citep{Shi2011,Tor2019}. 
But how the energy is accumulated in the corona and how flares are triggered are still not very clear. 
Generally speaking, magnetic field lines in the corona move in both random and organized behavior primarily because of dragging the field line footpoints by photospheric motions, which builds up a highly stressed magnetic field in the corona \citep{For2000, Che2011}.
Flares occur when magnetic energy in the corona accumulates to some extent and encounters some kind of triggering mechanism.
Shearing motion of the photosphere is believed to play a key role in powering active regions and even triggering flares, which creates a richness of magnetic shear in flare-prolific sites. \\

Over several decades, various trigger mechanisms have been proposed conceptually or with via magnetohydrodynamics (MHD) simulations to explain the concrete initiation processes of certain kinds of flares, all summarizing magnetic shear as the key feature from observations.
\citet{Moo2001} came up with the tether-cutting mechanism conceptually from sheared core magnetic field. 
That is, a filament is supported by dips of the magnetic field nearly aligned with a magnetic polarity inversion line (PIL), below which a current sheet is formed between the two oppositely curved elbow regions of the sheared core field. 
As the magnetic shear increases, magnetic reconnection commences, leading to a long concave-upward flux rope, which pulls up the filament to rise and make it erupt \citep{Son2021}. 
A resembling model was proposed by \citet{Van1989}, who pointed out that successive magnetic flux cancellation along the PIL of a sheared magnetic arcade would create the helical fields which can hold a filament. 
The helical rope could get higher and higher as flux cancellation goes on, which is quite an analogy to a filament eruption \citep{Ama2003}.
The flux cancellation could develop the twist of coronal fields, which can also cause filament eruption via MHD instability \citep{Sak1976,Tor2005,Fan2007,Dem2010}.
The so-called breakout model could be regarded as the external tether cutting mechanism \citep{Che2011}, where the upper part of the core magnetic region reconnects with the overlying magnetic field, which erases the downward magnetic tension force of the background field \citep{Ant1999,Aul2000,Arc2008}. 
All the triggering mechanisms mentioned above are made in the framework of a sheared magnetic field, which is supposed to be produced by photospheric shearing motion, especially sunspot shearing motion. 
However, to our knowledge, unambiguous evidence for the eruption of a flare driven by the direct shearing motion of a sunspot is still rare in literature.  \\

The solar lower atmosphere often has an immediate signature during the eruption of a sheared magnetic field, like unshearing motion in two-ribbon flares, often shown as the decrease of centroid distance of H$\alpha$ or hard X-ray kernels (so-called converging motion) during the rising phase of solar flares \citep{Ji2004, 2009ApJ...693..847L, Su2007a}.
In addition to the converging motion, there is a simultaneous decrease in the height of looptop sources or flaring loops, so the unshearing motion may also reflect the rapid shrinkage of the volume of the sheared magnetic field in the flare energy-releasing site \citep{2006ApJ...636L.173J, 2007ApJ...660..893J, 2013ApJ...777..152S, 2009ApJ...696..121L}. 
On the other hand, the shrinkage of the core magnetic field and the opening of the peripheral magnetic field, the restructuring or reconfiguration, will have an impact on the lower atmosphere.
One kind is the stepwise and permanent flare-related changes of photospheric magnetic fields or magnetic shear as reported by \citet{Wan1992} and \citet{Cam1999}.
Another is the back-reaction of flare to photosphere that seems to provide Lorentz force to it as firstly formulated by \cite{Hud2008}.
%Theoretically, \citet{Hud2008} assessed the enhancement of horizontal photospheric magnetic fields expected from the back-reaction of coronal magnetic reconfiguration by considering the vertical component of Lorentz force.
\citet{Fis2012} further formulated both horizontal and vertical Lorentz forces changes implied by the observed changes of vector magnetic fields related to flares and extended to discuss the back-reaction of photosphere under the condition of momentum and energy conservation. 
The force and the associated torque will change the state of photosphere horizontal motion, and thus the theory can be verified with observations. 
Several verifications have been made with the nice spatiotemporal correlation \citep{Wan2014, Bi2016},  and they are far from enough, we should seek further quantitative analysis.   \\ 

Here, we report an M3.3 flare powered and even triggered by pure sunspot shear motion started suddenly after a quasi-stationary state since it was formed.
The energy accumulation and triggering processes are distinct from the gradually long-term evolution process of active regions often shown as flux emergence/cancellation and/or photospheric motion such as the rotation of a sunspot or network magnetic fields \citep{Den2001, Tan2009, Kum2010, Sun2012, Vem2012, Liu2013, Bi2016, Yan2018, 2021ApJ...906...66D}. 
The sunspot was suddenly pulled back in the opposite direction immediately after about the flare's peak time, and, to our knowledge, this kind of rapid backward motion has not been reported before. 
The whole evolution is investigated from long-term aspects and we will also pay attention to rapid changes during the flaring period allowed by HMI and AIA data set.
We describe the data analysis in Section~\ref{sec:Data}. The results are represented in Section~\ref{sec:Res}.
Discussion and summary are given in Section~\ref{s:d&s}.

\begin{figure}
\includegraphics[width=18cm,clip=]{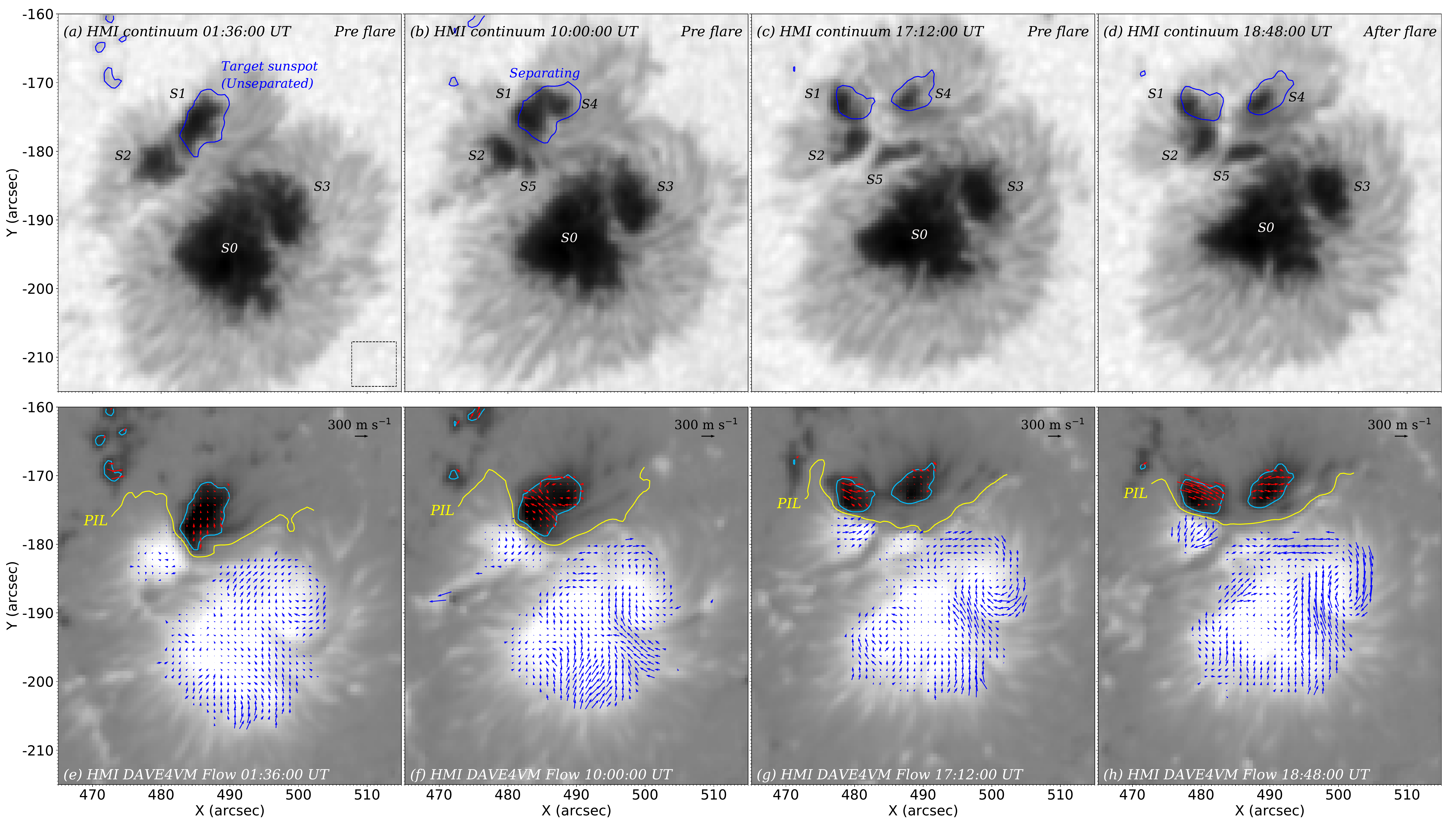}
\centering
\caption{
Top panels: Photospheric continuum emissions of AR 11818 observed by SDO/HMI at 01:36 UT, 10:00, 17:12, and 18:48 UT on August 17, respectively.
The sunspot we studied is labeled with ``S1'', which split out another sunspot labeled with ``S4'' at about 08:27 UT.
The main sunspot of the AR is labeled with ``S0''. 
Other sunspots are labeled with ``S2'', ``S3'' and ``S5''.
The sunspot S5 was separated from S2 at about the same period as S1 started to separate.
The region framed by the dotted rectangle in panel (a) is used in the calculation of the intensity of quiet region.
Bottom panels: The background maps are the vertical photospheric magnetic field ($B_z$) corresponding in temporal and spatial to the figures above. 
The arrows are the tracked horizontal local velocity vectors derived with the DAVE4VM method based on the data of the photospheric vector magnetic field \citep{Sch2008}. 
The blue and red are the tracking results for positive (negative) vertical magnetic field (B$_z$) greater (lower) than 1000 (-1000) G.
The length of red/blue arrows is made proportional to the horizontal local velocity.
The cyan contours are for B$_z$ = -1000 G and the yellow contours are for the PIL between the sunspots.
(An animation of this figure is available that presents the evolution of the AR in HMI LOS magnetograms and continuum intensities, and in AIA 131, 171, 304, and 1600 {\AA}.)
}
\label{fig1}
\end{figure}

%%%%%%%%%%%%%%%%%%%%%%%%%%%%%%%%%%%%%%%%%%%%%%%%%%%%%%%%%%%%%%%%%%%%%%%%
%%%%%%%%%%%%%%%%%%%%%%%%%%%%%%%%%%%%%%%%%%%%%%%%%%%%%%%%%%%%%%%%%%%%%%%%
%%%%%%%%%%%%%%%%%%%%%%%%%%%%%%%%%%%%%%%%%%%%%%%%%%%%%%%%%%%%%%%%%%%%%%%%
\section{Data analysis} \label{sec:Data}
The data used here were recorded by the Atmospheric Imaging Assembly \citep[AIA;][]{Lem2012} and the Helioseismic and Magnetic Imager \citep[HMI;][]{Sch2012} on board the Solar Dynamics Observatory \citep[SDO;][]{Pes2012}.
AIA takes full-disk EUV images in 94, 131, 171, 193, 211, 304 and 335 {\AA} with a time cadence of 12 s and UV images in 1600 and 1700 {\AA} of 24s. 
The photospheric line-of-sight (LOS) magnetograms and continuum intensities are obtained from HMI with a cadence of 45 s. 
HMI has also mapped the photospheric vector magnetic field every 720 s. 
For our research, the product of vector magnetograms was remapped to heliographic coordinates. 
All the images were reconstructed in a resolution of 0$\farcs$5.

\section{Results} \label{sec:Res}
The hosting active region NOAA 11818 on 2013 August 17, exhibiting a magnetic $\delta$ configuration, was located in the southwest of the solar disk (S07W30) when the M3.3 flare occurred. The flare started at 18:16 UT, peaked at 18:24 UT, and ended at 18:35 UT as observed in GOES 1-8 \AA\ flux curve.
Figure~\ref{fig1} presents four typical moments of the active region as observed with photospheric continuum intensities (top panels) and vertical magnetic field (B$_z$) (bottom panels).
The blue and red arrows in the lower panels represent the horizontal local velocity derived from the DAVE4VM method based on the magnetic features with positive (negative) B$_z$ greater (lower) than 1000 (-1000) G \citep{Sch2008}.
The target sunspot S1 appeared days before the flare at about 19:00 on August 15 and kept quasi-stationary relative to the main sunspot S0 of the AR until 08:27 UT on August 17. 
There existed other two sunspots S2 and S3 as shown in panels (a) and (e).
At about 08:27 UT on August 17, sunspot S1 was suddenly separated into two sunspots, and we label the sunspot in the east still with S1 and the west with S4.
The sunspot S2 also split into a small sunspot labeled with S5 at about the same period as displayed in panels (b).
Soon after, the target sunspot S1 started and kept a distinct northeastward motion along the PIL until about the peak time of the flare.
Meanwhile, sunspot S4 remained stationary relative to the main sunspot S0 as shown in panels (c) and (g).
When the flare came to its apex, i.e. about 18:24 UT, the moving direction of sunspot S1 underwent a sudden retraction from northeast to southwest, and at the same time, sunspot S4 had a sudden movement, which is displayed in panel (h).
An online animation (\textit{Fig1.mp4}) is available to show the evolution more vividly. \\

\begin{figure}
\includegraphics[width=18cm,clip=]{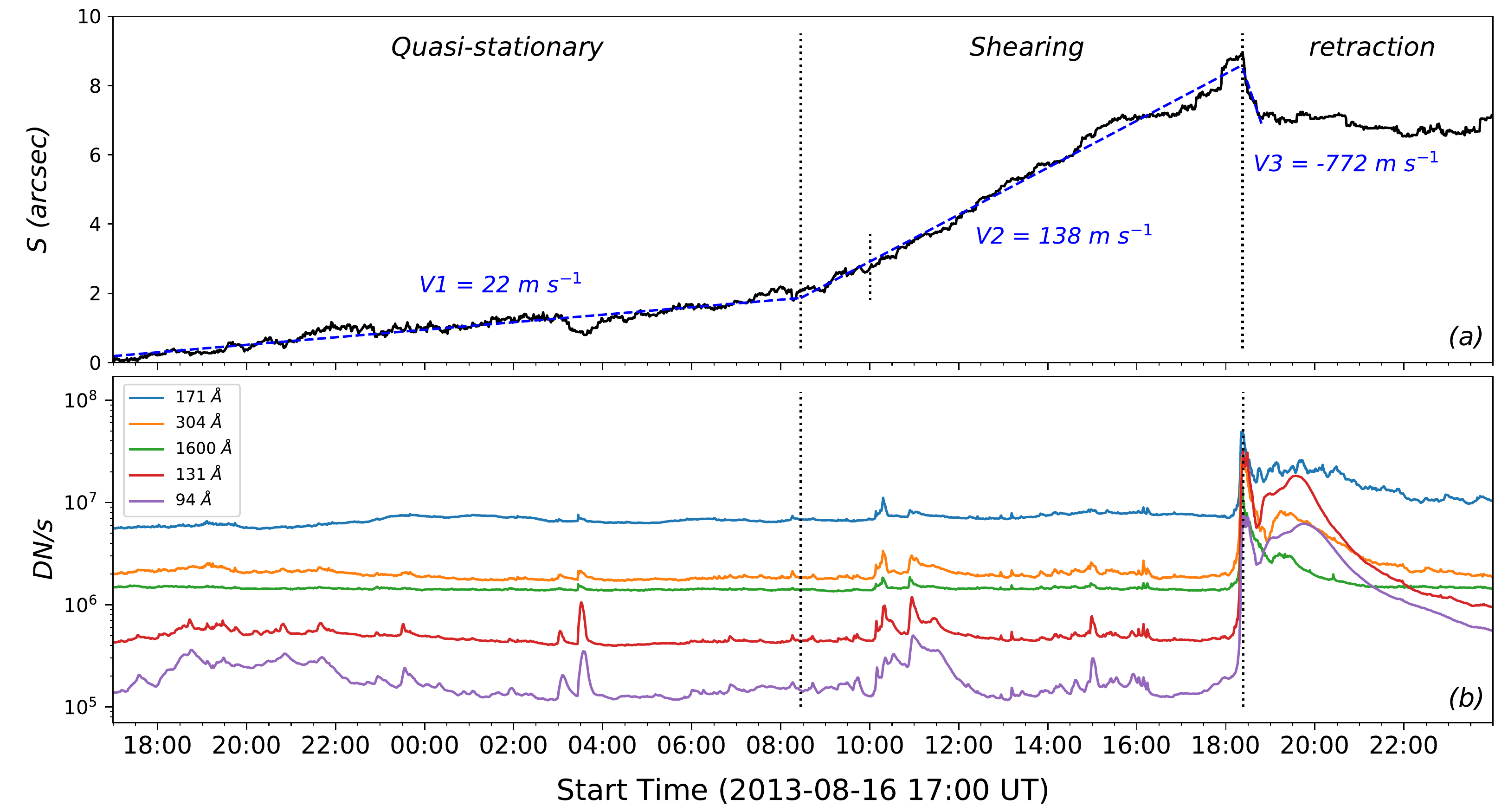}
\centering
\caption{(a) Time profile for the displacement of the target sunspot S1. 
The displacement was derived from the CoM calculation to the major part of the target sunspot smaller than 30000 DN/pixel in the continuum maps.
It should be noted that the curve is jointed by two curves on the left and right sides of 10:00 UT on August 17 (the junction position is pointed with a shorter dotted line) since,  at about that time, the target sunspot was separated from the initial one completely.
The linear fitting function for different moving speeds is drawn with dashed lines labeled with the corresponding speed value.
The longer vertical dotted lines separate the three phases.
(b) The emission flux variation of AIA 94, 131, 171, 304 and 1600 {\AA} of the AR. % with a field of view as the same as Figure~\ref{fig1}.
}
\label{fig2}
\end{figure}

\begin{figure}
\includegraphics[width=12cm,clip=]{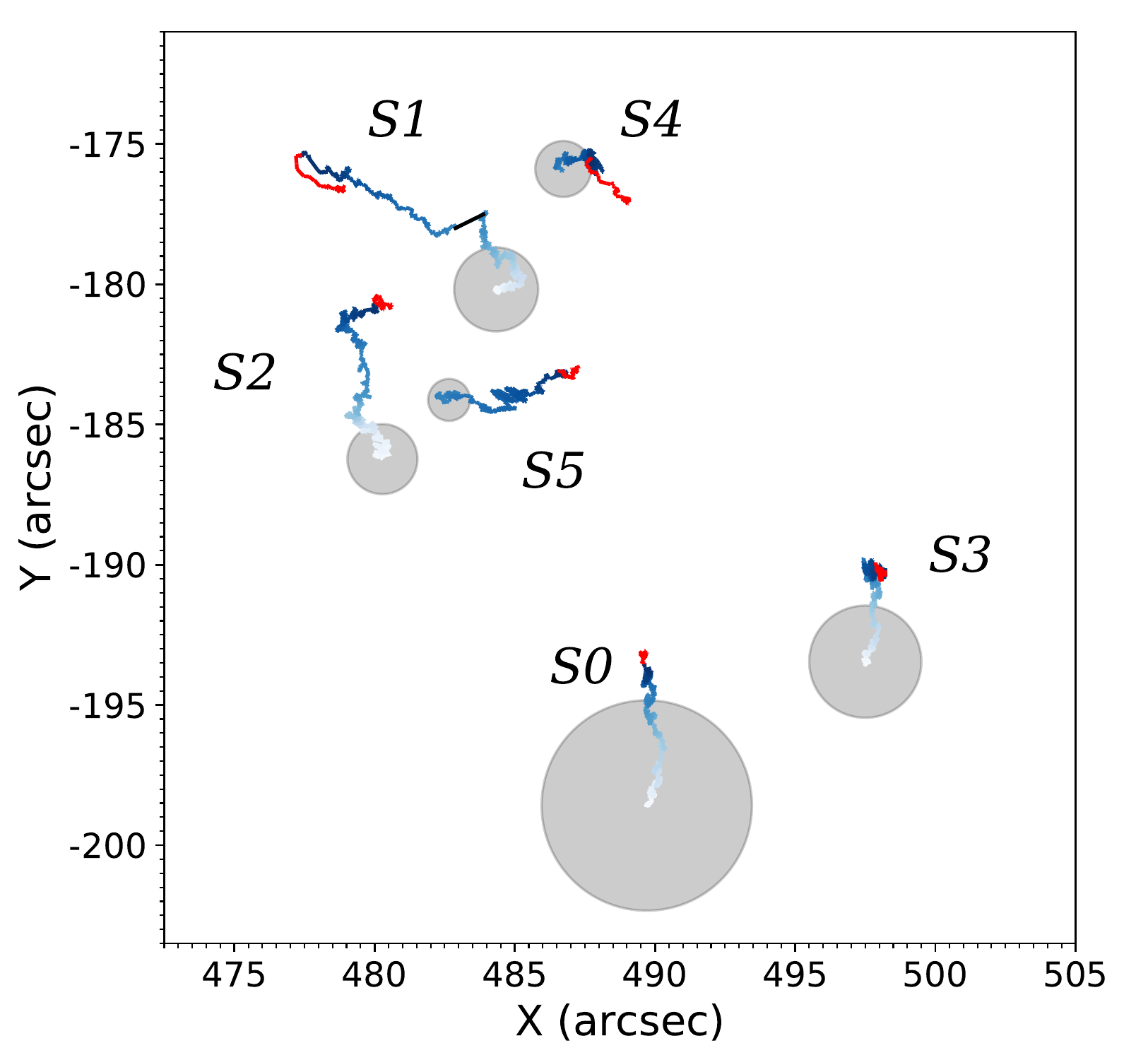}
\centering
\caption{The trajectories of sunspots S0, S1, S2, S3, S4 and S5 as shown in the upper panels of Figure~\ref{fig1}. 
The color varying from white to dark blue stands linearly for a different time from 17:00 UT on August 16 to the flare starting time.
The red curves of each sunspot exhibit their track during the flaring time until 19:30 UT on August 17.
The trajectories of sunspots S4 and S5 started at 10:00 UT on August 17.
The black segment in the track of sunspot S1 is due to S4 separating completely from S1 at about 10:00 UT on August 17, which is described in detail in the paper.
}
\label{fig3}
\end{figure}

To investigate the evolution of the target sunspot movement more comprehensively, we plot a time profile for the displacement of the target sunspot in Figure~\ref{fig2}(a) by using the center-of-mass (CoM) \citep{Anw1993} to estimate its position.
The x- and y-displacement of the sunspot at each moment in the CoM method are defined as 
\begin{equation} 
 X_d=\frac{\sum X_{i}I_{i}}{\sum I_{i}}-\frac{\sum X_{i0}I_{i0}}{\sum I_{i0}},Y_d=\frac{\sum Y_{i}I_{i}}{\sum I_{i}}-\frac{\sum Y_{i0}I_{i0}}{\sum I_{i0}}  
\label{form1}
\end{equation}

where X$_i$ and Y$_i$ are the x- and y-coordinate of each pixel in the sunspot at one moment and I$_i$ is the corresponding continuum density, and the subscript i0  represents the point of 17:00 UT on August 16 or 10:00 UT on August 17 respectively. We use a threshold of 52\% of quiet region (displayed by the dotted rectangle in Figure~\ref{fig1}(a)) intensity in the calculation.
The reason for the two starting points is that sunspot S4 was disjointed from S1 completely around 10:00 UT on August 17.
The AR had no detectable enhancement in HMI continuum intensity during the flare.
Therefore, there was no white-light emission that would bias the treatment of the sunspot position.\\

In this way, the centroid of the sunspot before and after this point of time was subtracted from its centroid at 17:00 UT on August 16 and 10:00 UT on August 17 respectively. The second time profile was shifted to connect the first time profile to form a complete one in order to demonstrate the whole moving process.
The connection is smooth due to the same slope. Clearly, the whole time profile is characterized by four stages with different speeds including the last horizontal one that is almost stationary.
Three time nodes are 08:27, 18:22, and 18:48 UT on August 17.
We mainly focus on the first three stages, which can be termed quasi-stationary, shearing, and retraction states. The moving pattern has been pictured by the red arrows in Figure~\ref{fig1}(e), (g), and (h), while those in Figure~\ref{fig1}(f) depict the separation process. The displacement of the first three stages is 2.1, 6.8, and -1.7$^{\prime \prime}$, with periods of about 15.5, 10.0, and 0.4 hours. 
Their linear speeds are 22, 138, and -772 m s$^{-1}$, respectively. As a comparison, the average speeds of the target sunspot represented with arrows in Figure~\ref{fig1}(e), (g), and (h) are calculated, which are 39, 107, and -182 m s$^{-1}$. It can be seen that the speed in the second stage is over 100 m s$^{-1}$ as derived from the two methods, which is regarded as the theoretical limit of shear flow that can lead to effective heating and cause flares \citep{Hey1984}. \\

The trajectories of all sunspots centroids obtained by the CoM are plotted in Figure~\ref{fig3}.
The portion of each curve that shades from white to dark blue represents the trajectory from 17:00 UT on August 16 to the flare starting time linearly, while the red portion outlines the track during the flaring time until 19:30 UT on August 17.
It should be noted that the starting color of the curve of sunspot S4/S5 is light blue rather than white, which is due to the centroid of the two sunspots being measured from 10:00 UT on August 17.
From the moving of sunspots S0, S2 and S3, We can clearly see that the whole active region had a general trend moving towards the north.
The movement of sunspots S1 and S4 were significantly affected by the flare while other sunspots were less affected as shown by the red trajectories.\\

Figure~\ref{fig2}(b) further presents the time profiles of AIA 94, 131, 171, 304 and 1600 {\AA} flux of the AR.
%are also been plotted in Figure~\ref{fig5}(b), which are obtained from the view field as the same as that of Figure~\ref{fig2}. 
Comparing the first two stages, we can clearly find out that the curves of the latter one contain more spikes.
It means that the active region has been effectively activated in this period.  All these spikes are confined eruptions that can be considered as a series of precursors of the M3.3 flare. \\

\begin{figure}
\includegraphics[width=18cm,clip=]{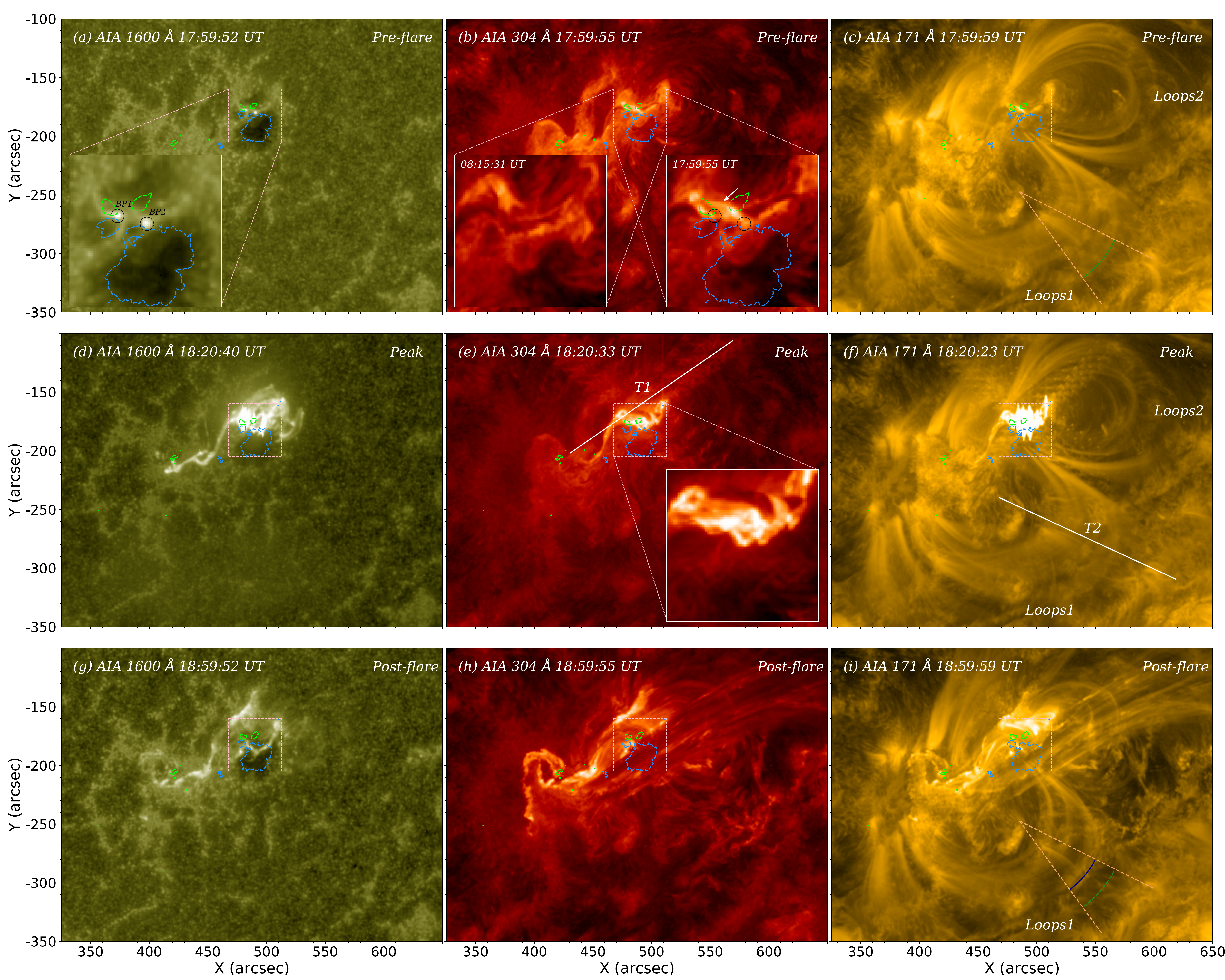}
\centering
\caption{Images of AIA 1600, 304 and 171 {\AA} overlaid with the contours of $\pm 1000$ G (blue/lime) of the vertical magnetic field at 18:00, 18:24 and 19:00 UT on August 17 in the first, second and third row, respectively. 
Top panels: Snapshots about 24 minutes before the peak time of the flare. 
The sub-figure in panel (a) shows an enlarged view of the AIA 1600 {\AA} image in the boxed region.
The dashed circles labeled with BP1 and BP2 show the two brightening points in the image. 
The sub-figure in panel (b) displays the enlarged view of the AIA 304 {\AA} image with the same boxed region in panel (a) at 08:15 UT and 17:59 UT on August 17, respectively. 
The arrow in the right sub-figure points to the braided structure in the filament.
The two dashed circles are the same as the circles in panel (a). 
The two coronal loop systems are indicated with the annotation ``Loops1'' and ``Loops2'' in panel (c). 
Middle panels: Snapshots about 4 minutes before the peak time of the flare. 
The slits T1 and T2 in panels (e) and (f) are used to investigate the motion of the filament and Loops1, respectively. 
Bottom panels: Snapshots about 36 minutes after the peak time of the flare.
Auxiliary lines plotted near the loop apexes on panels (c) and (i) exhibit the contraction motion.
}
\label{fig4}
\end{figure}

\begin{figure}
\includegraphics[width=18cm,clip=]{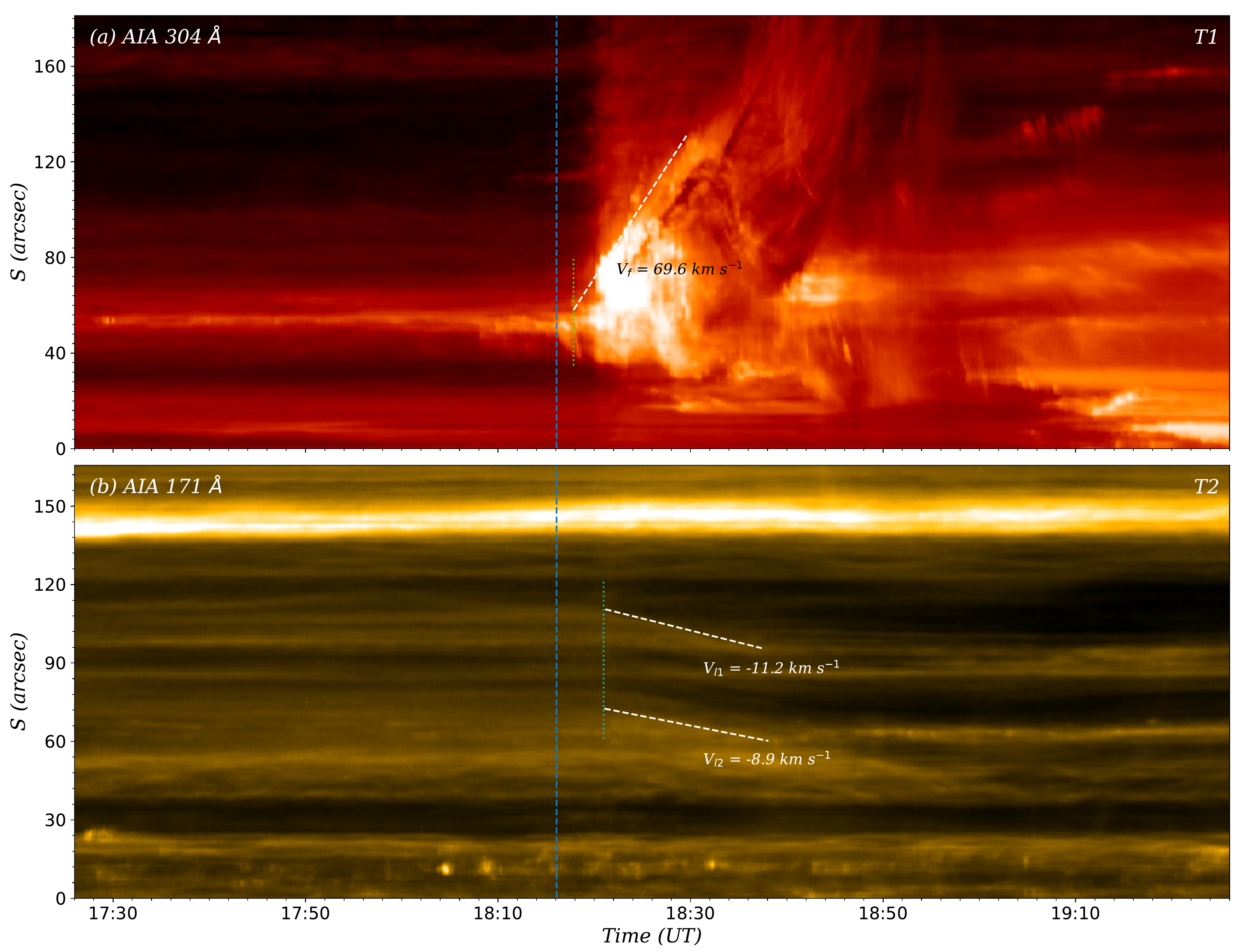}
\centering
\caption{Time-distance diagrams of the slits T1 and T2 shown in Figure~\ref{fig2}(e) and (f).
The blue dashed line signifies the start time of the flare, and the launching time of the filament eruption and the contraction of Loops1 are delineated by the two green dotted lines respectively. 
The white dashed lines represent the linear fitting for the motion of the filament and Loops1.}
\label{fig5}
\end{figure}

\begin{figure}
\includegraphics[width=18cm,clip=]{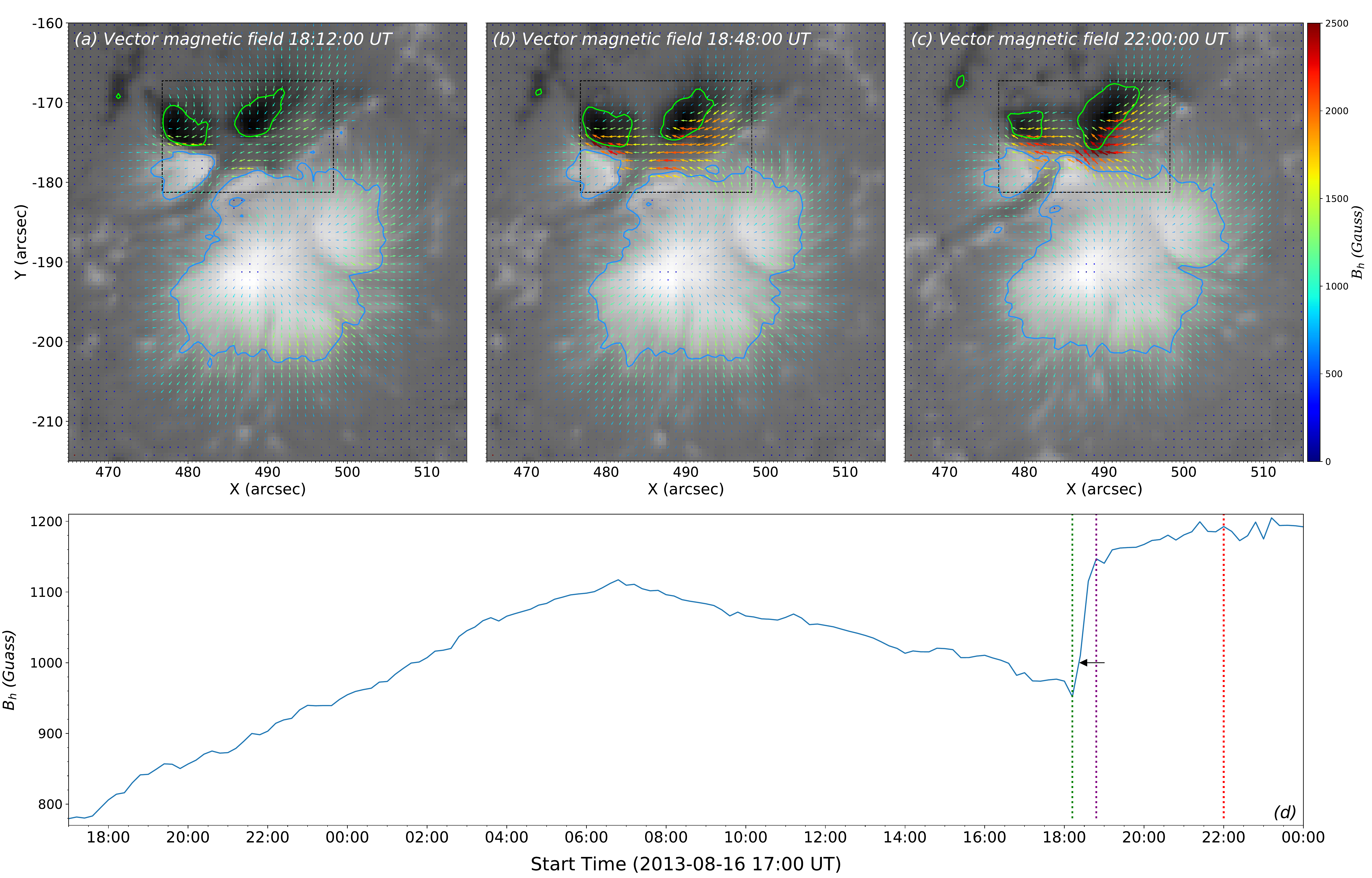}
\centering
\caption{Top panels: The background maps are the vertical photospheric magnetic field ($B_z$) while the arrows represent the horizontal magnetic field ($B_h$). The blue/lime contours are for $\pm 1000$ G of $B_z$.
Bottom panel: The mean $B_h$ variation calculated from the region framed with a black dashed rectangle in the top panels.
The three dotted lines correspond to the moments of three figures in the upper panels from left to right.
The arrow in panel (d) refers to the time when the moving direction of sunspot S1 was reversed, that is 18:22 UT, as is shown in Figure~\ref{fig2}.}
\label{fig6}
\end{figure}

Like most M- and X-class flares \citep{Yan2011,Zha2022}, the M3.3 flare was also associated with a filament eruption as shown in Figure~\ref{fig4}. A bright twisting or braided structure appeared on the central part of the filament around 18:00 UT, which can be seen clearly in the right sub-figure of panel (b). At the same time, two brightening points (BP1 and BP2 in panel (a)) showed up in the chromosphere just below the center part of the filament, and one of them (BP1) was in contact with sunspot S1. The consistency in temporal and spatial between BP1, BP2 in the chromosphere and the braided structure on the filament hints that these two kinds of structures are the manifestations in different heights of just the one magnetic reconnection process. The spatial overlap between sunspot S1 and BP1 suggests that the magnetic fields of S1 might be involved in the reconnection. Starting at about 18:16 UT, the filament was activated and partially erupted out entangled with the braided structure, eventually forming a two-ribbon flare as shown in panels (e) and (h). \\

After filament eruption, there appeared a contraction of side loops. We present AIA 171 {\AA} observations for this event in the third column of Figure~\ref{fig4}.
Two coronal loop systems (Loops1 and Loops2) are visible in the south and northwest of the filament before the flare, as is exhibited in panel (c).
At about 18:21 UT, the Loops1 suddenly moves inward to the center of the flare core region as shown by the auxiliary lines plotted in panels (c) and (i), while the majority of Loops2 burst out with the filament. 
The contraction lasted about 20 minutes, even throughout the flare decaying phase. 
It should be noted that the reversal of the moving direction of the target sunspot also occurred at this time. 
Two slits are placed across the Loops1 and the filament to construct time-distance diagrams (Figure~\ref{fig5}). 
The inward motion shows an apparent speed of about ten kilometers per second and the projected speed of filament is on the order of tens of kilometers per second. 
The starting time of filament eruption (about 18:18 UT) was about three minutes ahead of that of the contraction of Loops1 (18:21 UT), which is consistent with the theoretical prediction that a smaller magnetic pressure gradient will be caused by the depletion of magnetic energy, and overlying magnetic fields must contrast to strike a new balance. \\

In addition, a sudden and permanent enhancement of the photospheric horizontal field ($B_h$) ensued from the beginning of the flare at 18:12 UT.
In the top panels of Figure~\ref{fig6}, a core region (enclosed by the dashed rectangle) along the PIL is identified to show the noticeable change before, during and after the flare. 
The background maps are the vertical photospheric magnetic field ($B_z$) while $B_h$ is highlighted by the arrows aligned to the field direction, with varying colors proportional to the different field strengths as manifested by the color bar. 
Furthermore, the time profile of average $B_h$, calculated from the dashed rectangle region, is plotted in panel (d).
Before the flare, the strengths experienced an increment of 340 G in 14 hours and a reduction of 165 G in 11 hours, which should belong to the gradual evolution of the photospheric magnetic field. 
And then a sharp increment of 230 G can be observed in 36 minutes during the flaring time, with the change-over cotemporal with the impulsive phase of AIA curves in Figure~\ref{fig2}(b). 
This suggests a more horizontal photospheric magnetic field around the PIL after the flare. 
Considering the coronal loop contraction, both pictures endorse the magnetic implosion scenario \citep{Hud2000,Hud2008,Wan2010,Fis2012}. \\

\begin{figure}
\includegraphics[width=18cm,clip=]{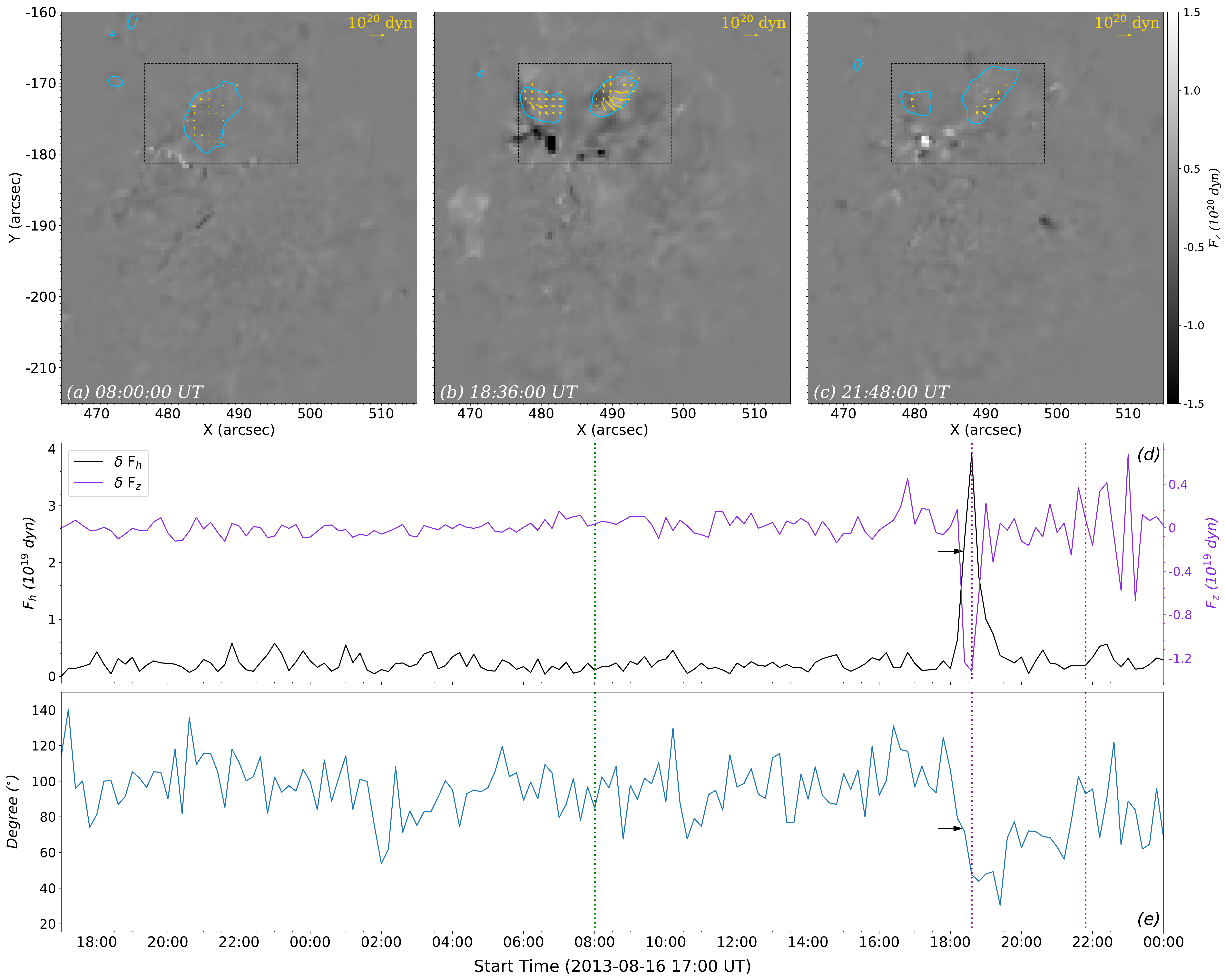}
\centering
\caption{Top panels: the background maps are the running differential vertical Lorentz force ($\delta F_z$) maps while the yellow arrows represent the running differential horizontal Lorentz force ($\delta F_h$) imposed on S1 and S2 enclosed by the blue contours, which are for -1000~G of $B_z$. 
Middle panel: the black/violet curve is the time profile of mean $\delta F_h$ imposed on S1 and S2/mean $\delta F_z$ in the region framed by the dashed rectangle shown in the upper panels, which is the same as that in the upper panels of Figure~\ref{fig5}. 
Bottom panel: the time profile of the mean angle formed between the horizontal local Lorentz force and the horizontal local velocity, which is obtained in the region enclosed by the blue contours in the upper panels.
The three dotted lines correspond to the moments of three figures in the upper panels from left to right.
The arrows in panels (d) and (e) refer to the time when the moving direction of sunspot S1 was reversed, that is 18:22 UT, as is shown in Figure~\ref{fig2}.
}
\label{fig7}
\end{figure}

To investigate the possible reasons for the response of the photosphere to the flare, the Lorentz force changes on the photosphere are further derived by applying the method proposed by \citet{Fis2012}. 
The background maps of the top panels of Figure~\ref{fig7} are the running difference maps of vertical Lorentz force ($\delta F_z$) with the dark/bright features exhibiting the force directed downward/upward. 
The running difference horizontal Lorentz force ($\delta F_h$) is highlighted by the golden arrows in the contours for $B_z$ = -1000 G, with varying lengths proportional to the different force strengths. 
Panel (d) presents the time profiles of the mean $\delta F_h$ in the contour and $\delta F_z$ in the dotted rectangle, which is the same as that in Figure~\ref{fig6}.
Obviously, both $\delta F_h$ and $\delta F_z$ are nearly zero at most times, except for an impulse signal during the flaring time, with the values of 3.9 $\times$ 10$^{19}$ dynes for $\delta F_h$ and -1.3 $\times$ 10$^{19}$ dynes for $\delta F_z$. 
It means that F$_z$, which should be interpreted as the magnetic pressure, got strengthened and resulted in the enhancement of B$_h$.
It stands for the downward collapse to the photospheric flare core region of magnetic fields, contrasting with the upward component of the flare according to the magnetic implosion idea \citep{Hud2000,Fis2012}.
On the other hand, F$_h$ was also strengthened mainly on sunspots S1 and S4 along the direction of retraction motion, that is, the direction toward the core region.
Similarly, it contrasts with the laterally outward component of the flare considering the conservation of momentum in the horizontal direction. 
Both the change of F$_z$ and F$_h$ during the flaring time depicts the process in which the magnetic fields erupt outward and then collapse inward.\\

To support the picture, the bottom panel of Figure~\ref{fig7} gives the time profile for the mean angle formed between the horizontal Lorentz force and the horizontal velocity field, averaged in the region enclosed by the blue contours in the upper panels of Figure~\ref{fig7}.
We see that there is a sudden drop for the mean angle, showing that the horizontal Lorentz force was pulling the sunspot backward during the flare period. 
The tightly temporal and spatial correlations between the above variations and flare eruption, strongly imply that the retraction motion of sunspot could result from Lorentz force change induced by the flare \citep{Fis2012}. \\

For an order-of-magnitude estimation, we compare the momentum change of the sunspots S1 and S4 with the impulse of F$_h$ during the rapid change in sunspot velocity. 
The linear fitting results 138 m s$^{-1}$ and -772 m s$^{-1}$ are used as the velocity at the starting and ending times of the period, respectively. 
The average F$_h$ in flaring time,  $\sim$ $ 2.5 \times 10^{22}$ dyne, is taken as the average F$_h$ in this process. 
The surface area of the sunspots S1 and S4 ($B_z$ smaller than -1000 G) is about $2.2 \times 10^{7}$ Mm$^2$, and the density is taken as $4 \times 10^{-7}$ g cm$^{-3}$ \citep{Wan2014}. 
Under the assumption that the sunspot is a thin flux tube \citep{Spr1974,Spr1981},  depth is take as 1.0 Mm \citep{Anw1993,Bi2016}. 
Then the mass of sunspots S1 and S4 is estimated to be $\sim 8.8 \times 10^{15}$ kg, and the change of the sunspots momentum is about $8.0 \times 10^{18}$ kg m s$^{-1}$.  
According to momentum conservation, these figures lead to the estimates of the timescale of $\sim$ 32 s. 
If the viscosity is taken into account, the timescale will be longer, possibly up to several minutes, which is consistent with our observations and previous reports \citep{Anw1993,Wan2014,Bi2016}.

\section{Discussion and Summary} \label{s:d&s}

In this Letter, we give a detailed analysis of the M3.3 flare of  August 17, 2013 (SOL2013-08-17T18:16), showing the well-defined associations between the rapid sunspot shearing motion of the small sunspot and the activation of the filament tied to the sunspot before the flare, between the sudden retraction motion of the same sunspot and the deduced Lorentz force from the rapid reconfiguration of the coronal magnetic field upon the flare onset. \\
 
By using HMI intensity images and vector magnetograms, we find that the flare was powered and triggered by a rapid sunspot shear motion that lasts $\sim$10 hours until the flare onset. Compared with previous reports in the literature that the photospheric shearing motions were generally characterized by the long-term gradually evolution along PILs and were often accompanied by other photospheric phenomena like flux emergence/cancellation, sunspot rotation, etc \citep{Den2001, Yan2004, Kum2010, Sun2012, Vem2012, Liu2013, Yan2018}, the shearing motion in this event and its association with flare are simply and neat as observed from the small moving sunspot. It started its shearing motion abruptly from the quasi-stationary state and underwent an abrupt retraction motion when the flare started. Though it seems that pure shear motion can not provide the whole energy for filament eruption/flare in a limited time from a potential state \citep{Che2000}.
In this case, AR 11818, might have already accumulated plenty of free magnetic energy, a further input will occur as the sunspot kept its shearing motion along the PIL.
The shear motion was followed by an increased number of brightening spikes noticeably in AIA flux curves, indicating the contributing processes of magnetic energy to the corona. The contribution is made via either the formation of sheared magnetic configuration resembling the process proposed by \citet{Van1989} or the formation of unstable hot channels as observed by \citet{Wan2018}. In this event, the shearing motion can be linked with the twisted magnetic structure of the filament.
Therefore, the picture is similar to the shear reconnection model proposed by \citet{Van1989}, where shear flow along with the PIL transfer the arcade field in corona to the helical field by reconnection, which is capable of holding a filament. As the reconnection goes on, the magnetic flux in the helical field becomes too large to keep its equilibrium, which leads to the helical field erupting out with the filament in it \citep{Sun2012}. \\

The more important and interesting observation is that the small sunspot S1 underwent a sudden retraction motion with a much larger speed after the impulsive phase of the flare. Because the time for the direction reversal is too short for a turnover of the photospheric convection \citep{Bi2016}. The phenomena can be explained with the Lorentz force produced during ``magnetic implosion'' \citep{Hud2000, Hud2008, Wan2010, Fis2012} under the condition of momentum and energy conservation. It is shown that the direction of the impulse of $\delta F_h$ in the flaring time is consistent with that of sunspot reversal, we can believe that the sudden retraction motion results from the horizontal Lorentz force produced by the coronal magnetic field reconfiguration rather than photospheric convective flows.
The back-reaction on the photospheric shear motion has various forms \citep{Tan2009, Wan2014, Liu2016, Xu2017}, but none of them showed a 180$^{\circ}$ change of direction. There still existed vertical (downward) Lorentz force exerted on the photosphere as manifested as the rapid and irreversible increase of the photospheric horizontal magnetic field. It should be noted that the two patterns of Lorentz force play different roles in the photosphere complex. That is, vertical (downward) Lorentz force acts to increase the photospheric shear while horizontal to decrease the shear \citep{Pet2013}.  \\

In summary, the M3.3 solar flare in AR 11818 on August 17, 2013, is a rare valuable event, regarding the sunspot motions before, during, and after the flare. 
Thus, it can serve as a good example of showing the magnetic coupling across different layers of the solar atmosphere during a solar flare. 
It provides direct evidence that the shearing motion of a sunspot can power and trigger a solar flare. 
It especially confirms that the fast relaxation or restructuring of coronal fields can have an impact on the photosphere, resulting in sudden Lorentz force as well as an enhanced horizontal magnetic field.  
Magnetic coupling processes across the solar atmosphere in other flares may take different forms, like the rotation of sunspots or network magnetic fields,  the emergence, cancellation, and rapid changes of magnetic fields, but the signature is vague in most flares. 
This may be due to insufficient spatiotemporal resolution and low accurate polarimetry \citep{2003SPIE.4853..240K, 2020SoPh..295..172R}.  Further similar systematic research with the aid of well-observed cases will provide further insight into the coupling between photosphere and corona during solar flares.

\begin{acknowledgements}

The authors are grateful to the anonymous referee and colleagues in Purple Mountain Observatory and Yunnan Observatories for their constructive suggestions and comments.
SDO is a mission of NASA\rq{}s Living With a Star Program. AIA and HMI data are courtesy of the NASA/SDO science teams. 
This work is supported by the National Key R\&D Program of China 2021 YFA1600500(2021YFA1600502), the NSFC grants (No. 11790302, 11790300, 12173084, 12073072), CAS Key Laboratory of Solar Activity, and the Strategic Priority Research Program on Space Science, CAS (XDA15052200, XDA15320301), China Postdoctoral Science Foundation 2020M671639, Yunnan Key Laboratory of Solar Physics and Space Exploration 202205AG070009.
\end{acknowledgements}

\bibliography{bibtex}{}
\bibliographystyle{aasjournal}

\end{document}